\documentclass{article} %
\usepackage{iclr2023_conference,times}

\usepackage{amsmath,amsfonts,bm}

\def\eqref#1{equation~\ref{#1}}

\def\1{\bm{1}}

\DeclareMathAlphabet{\mathsfit}{\encodingdefault}{\sfdefault}{m}{sl}
\SetMathAlphabet{\mathsfit}{bold}{\encodingdefault}{\sfdefault}{bx}{n}

\usepackage[hidelinks]{hyperref}
\usepackage{url}
\usepackage{xspace}
\usepackage{graphicx}
\usepackage{tikz}
\usepackage{booktabs}
\usepackage{multirow}
\usepackage{caption}
\usepackage{multicol}
\usepackage{tablefootnote}
\usepackage{threeparttable}
\usepackage{makecell}
\usepackage{enumitem}

\title{Conversational Automated Program Repair}

\author{Chunqiu Steven Xia, Lingming Zhang\\
University of Illinois at Urbana-Champaign\\
\texttt{\{chunqiu2, lingming\}@illinois.edu} \\
}

\iclrfinalcopy %
\begin{document}

\newcommand{\CodeIn}[1]{{\small \texttt{#1}}}
\newcommand{\lingming}[1]{\textcolor{red}{Lingming: #1}}
\newcommand{\steven}[1]{\textcolor{blue}{Steven: #1}}
\newcommand{\mypara}[1]{\vspace{.03in}\noindent \textbf{#1}}
\newcommand{\Contrib}{$\star$\xspace}
\newcommand{\Comment}[1]{}
\newcommand{\STAB}[1]{\begin{tabular}{@{}c@{}}#1\end{tabular}}

\newcommand{\alpharepair}{AlphaRepair\xspace} 
\newcommand{\codebert}{CodeBERT\xspace}

\newcommand{\rewardrepair}{RewardRepair\xspace}
\newcommand{\recoder}{Recoder\xspace}
\newcommand{\deepdebug}{DeepDebug\xspace}
\newcommand{\cure}{CURE\xspace}
\newcommand{\coconut}{CoCoNuT\xspace}
\newcommand{\dlfix}{DLFix\xspace}
\newcommand{\sequencer}{SequenceR\xspace}
\newcommand{\tbar}{TBar\xspace}
\newcommand{\prapr}{PraPR\xspace}
\newcommand{\avatar}{AVATAR\xspace}
\newcommand{\simfix}{SimFix\xspace}
\newcommand{\fixminer}{FixMiner\xspace}
\newcommand{\capgen}{CapGen\xspace}
\newcommand{\jaid}{JAID\xspace}
\newcommand{\sketchfix}{SketchFix\xspace}
\newcommand{\nopol}{NOPOL\xspace}
\newcommand{\jgenprog}{jGenProg\xspace}
\newcommand{\jmutrepair}{jMutRepair\xspace}
\newcommand{\jkali}{jKali\xspace}
\newcommand{\genprog}{GenProg\xspace}
\newcommand{\angelix}{Angelix\xspace}
\newcommand{\spr}{SPR\xspace}
\newcommand{\aande}{AE\xspace}
\newcommand\arja{ARJA\xspace}
\newcommand\kalia{Kali-A\xspace}
\newcommand\card{Cardumen\xspace}
\newcommand\npefix{NPEFix\xspace}
\newcommand\gproga{GenProg-A\xspace}
\newcommand\rsrepair{RSRepair-A\xspace}
\newcommand\dynamoth{DynaMoth\xspace}

\newcommand{\codex}{Codex\xspace}
\newcommand{\codegen}{\textsc{CodeGen}\xspace}
\newcommand{\codegenmono}{\textsc{CodeGen-Mono}\xspace}
\newcommand{\codegenmulti}{\textsc{CodeGen-Multi}\xspace}
\newcommand{\chatgpt}{ChatGPT\xspace}
\newcommand{\gptj}{GPT-J\xspace}
\newcommand{\gptneo}{GPT-Neo\xspace}
\newcommand{\gptneox}{GPT-NeoX\xspace}
\newcommand{\incoder}{\textsc{InCoder}\xspace}
\newcommand{\codet}{CodeT5\xspace}

\newcommand{\llm}{LLM\xspace}
\newcommand{\llmfull}{Large Pre-Trained Language Model\xspace}
\newcommand{\clmfull}{Generative Model\xspace}
\newcommand{\clm}{generative model\xspace}
\newcommand{\apr}{APR\xspace}
\newcommand{\aprfull}{Automated Program Repair\xspace}
\newcommand{\nmt}{NMT\xspace}
\newcommand{\nmtfull}{Neural Machine Translation\xspace}
\newcommand{\nlp}{NLP\xspace}
\newcommand{\nlpfull}{Natural Language Processing\xspace}
\newcommand{\mlm}{infilling model\xspace}
\newcommand{\mlmfull}{Infilling Model\xspace}
\newcommand{\rtd}{RTD\xspace}
\newcommand{\rtdfull}{Replaced Token Detection\xspace}

\newcommand{\dfj}{Defects4J\xspace}
\newcommand{\quix}{QuixBugs\xspace}
\newcommand{\bears}{Bears\xspace}
\newcommand{\bip}{BugsInPy\xspace}
\newcommand{\manybugs}{ManyBugs\xspace}

\newcommand{\learning}{learning-based\xspace}
\newcommand{\Learning}{Learning-based\xspace} %
\newcommand{\template}{template-based\xspace}
\newcommand{\heuristic}{heuristic-based\xspace}
\newcommand{\constraint}{constraint-based\xspace}

\newcommand*\circled[1]{\scalebox{0.8}{\tikz[baseline=(char.base)]{
\node[anchor=text, shape=circle,fill, inner sep=0pt, minimum size=1.2em] (char) {\footnotesize \textcolor{white}{#1}};}}}

\captionsetup[figure]{font=bf,skip=2pt}%
\captionsetup[table]{font=bf,skip=2pt}%
\newcommand{\distance}{5pt}
\setlength{\textfloatsep}{1pt}%
\setlength{\floatsep}{\distance}%
\setlength{\intextsep}{\distance}%
\setlength{\dbltextfloatsep}{\distance} %
\setlength{\dblfloatsep}{\distance} %

\maketitle

\begin{abstract}

\aprfull (\apr) can help developers automatically generate patches for bugs. Due to the impressive performance obtained using \llmfull{s} (\llm{s}) on many code related tasks, researchers have started to directly use \llm{s} for \apr. However, prior approaches simply repeatedly sample the \llm given the same constructed input/prompt created from the original buggy code, which not only leads to generating the same incorrect patches repeatedly but also miss the critical information in testcases. To address these limitations, we propose \emph{conversational \apr}, a new paradigm for program repair that alternates between patch generation and validation in a conversational manner. In conversational \apr, we iteratively build the input to the model by combining previously generated patches with validation feedback. As such, we leverage the long-term context window of \llm{s} to not only avoid generating previously incorrect patches but also incorporate validation feedback to help the model understand the semantic meaning of the program under test. We evaluate 10 different \llm including the newly developed \chatgpt model to demonstrate the improvement of conversational \apr over the prior \llm for \apr approach. 

\end{abstract}

\section{Introduction}

Bugs in software can cause significant financial losses~\cite{bug_loss} and create dangerous health and safety problems~\cite{bug_safety}. Due to the high manual cost of fixing bugs~\cite{debuggingtime}, \aprfull (\apr)~\cite{gazzola2019aprsurvey} is a promising solution to reduce developer work by automatically generating patches given the buggy code and failing testcases.

Traditionally, \apr approaches commonly use the paradigm of Generate and Validate (G\&V), where \apr tools will first generate a list of candidate patches given the original buggy code and then validate each one sequentially until a \emph{plausible patch} that passes all the testcases is found. Plausible patch is then passed on to a human developer where they have to determine if this is a \emph{correct patch} that correctly fixes the underlying bug. Traditional \apr approaches such as template-based tools~\cite{ghanbari2019prapr, liu2019tbar, lou2020can} have been proven useful in fixing bugs with pre-defined templates to match buggy and corresponding fix code patterns. Recently, researchers have designed \learning \apr tools~\cite{ye2022rewardrepair, zhu2021recoder, jiang2021cure} which build a \nmtfull (\nmt) model by training on pairs of buggy and patch code. However, these \learning \apr tools suffer from lack of patch variety as it can only repair the types of bugs that are a part of the buggy/patch training data. Furthermore, these bug fixing datasets can be difficult to construct as it require scraping open-source bug fix commits which may contain many false positives, adding noise to the dataset. 

Recognizing the limitation of prior \learning \apr tools, researchers have started to look into directly leveraging \llmfull{s} (\llm{s}) for \apr without fine-tuning. \llm{s} have proven their ability in various code generation tasks~\cite{austin2021synthesis}. \cite{xia2022alpharepair} first introduced \emph{cloze-style} \apr where a \llm directly fill-in the correct code given its surrounding context. Other studies~\cite{prenner2021codexws, kolak2022patch, xia2022repairstudy} have also investigated directly applying different types of \llm{s} for \apr by smartly applying prompts or giving original buggy code as context. Typically, directly applying \llm{s} for \apr involves creating a common prompt/prefix which can be just the buggy context (zero-shot) or combining buggy context with a few examples of bug fixes (few-shot) as input to the model. Following the G\&V paradigm, prior approach will sample the \llm{s} multiple times to obtain candidate patches. However, this pipeline has the following limitations: 

First, sampling from the same prefix/prompt multiple times can lead to many repeated patches due to the probabilistic nature of sampling. This means the \llm{s} may waste a lot of compute and time generating the same patches which have already been validated as incorrect by the testsuite. Second, prompts provided to the \llm{s} for \apr are created only from the original buggy code and does not include any of the testcase information. Such information like the expected input and output examples that can help \llm{s} understand the functionality of the buggy program are not provided. Third, prior approaches also fail to consider the outputs produced by the generated incorrect patches. Previously incorrect patches may fail on a particular corner case, which can be exposed by looking at the test output and providing it to the \llm to address it in future patches.  

\mypara{Our Work.} We propose \emph{conversational \apr{}} -- a new paradigm of using \llm{s} for \apr that directly leverages the testcase validation information to provide feedback to \llm{s} in a conversational manner. In conversational \apr, we interleave patch generation with validation where \llm first generates a patch, we then validate it against testsuite to provide feedback and prompt \llm with the new feedback information to generate a new patch. While in this paper we consider simple testcase input/output/error validation feedback, one can apply conversational \apr with a wild range of possible feedback information such as human evaluation of the patch. We refer to the process of generating a patch followed by validation as a \emph{turn} where a conversation \emph{chain} is made up of multiple turns in sequence. In the start of the conversation chain, we begin with an initial prompt and sample the \llm to obtain a candidate patch. As we continue the conversation, the input given to the \llm in each turn is a concatenation of all previously incorrect patches along with their associated testcase feedback within the same conversation chain. A conversational chain is terminated once a patch that passes all the testcases are found or the maximum chain length is reached (i.e., maximum number of turns). In the latter case, we start a new conversation chain with the initial prompt again.

Compared with prior \llm for \apr tools which only use the buggy code snippet as inputs, conversational \apr incorporates patch validation in the form of validation feedback to help the model understand the \textit{reason} why previously generated patches are incorrect. Such feedback can contain the incorrect and expected test outputs or indicate if the generated patch contains compilation/runtime errors. Furthermore, while prior \llm for \apr tools continuously sample from the same input, our approach iteratively builds the input by including previously incorrect patches. As such, the \llm, through its long context window, can recognize previous generations and avoid repeatedly generating an already validated incorrect patch. We evaluated our conversational \apr by using 10 popular \llm{s}, where we found that our approach not only improves the number of bugs fixed but also can arrive at the correct patch faster compared with sampling-based baseline. Furthermore, we also evaluate the recently developed \chatgpt~\cite{chatgpt}\footnote{While we perform repair using \chatgpt, no part of this paper is written by \chatgpt. :)}, a dialogue focused \llm trained using reinforcement learning and highlight the performance of conversational \apr when using a \llm designed for conversation/dialogue.

\section{Background \& Related Work}

\subsection{\llm{s} for \apr}

To combat the reliance on training using bug-fixing datasets to build learning-based \apr tools based on \nmt models, researchers directly applied \llm{s} for \apr without any fine-tuning. \cite{xia2022alpharepair} proposed \alpharepair, the first \textit{cloze-style} \apr to directly leverage \llm{s} for \apr in a zero-shot setting by removing the buggy line and replacing it with masked tokens. \alpharepair then queries the \codebert~\cite{feng2020codebert} model to fill-in the masked tokens with the correct tokens to generate patches. \cite{prenner2021codexws} investigated the ability for \codex~\cite{chen2021codex} to repair bugs using a simple prompting method to generate a complete patched function given the original buggy function. \cite{kolak2022patch} evaluated the scaling effect of \llm{s} for \apr by using 4 \llm{s} of different model sizes to generate a single line fix given only the original buggy prefix (i.e., removing all lines after and including the buggy line of the buggy function). Recently, \cite{xia2022repairstudy} conducted an extensive study on directly applying \llm{s} for \apr. In the study, they adopt several repair settings, including few-shot generation using a few examples of bug fixes, cloze-style \apr and also single line generation. 

The findings across these prior work is consistent in showing that directly using \llm{s} for \apr achieves comparable if not better performance compared to prior \apr tools. However, these proposed \llm{s} for \apr techniques almost exclusively use sampling where patches are generated by sampling from the same input over and over again, leading to many repeated patches. Furthermore, the inputs to the \llm{s} are only constructed from the original buggy function, missing the rich information in the form of testcases. In this work, our conversational \apr approach aims to bridge these limitations in \llm{s} for \apr by constructing new inputs based on prior incorrect patches to avoid sampling repeated patches and providing the validation feedback to add another dimension of input apart from original buggy code to help the model understand the semantic meaning of the program. 

\subsection{Multi-Step Program Reasoning and Synthesis using \llm{s}} 

A related research direction is in applying multi-step reasoning for code understanding and synthesis.  
\cite{nye2021scratchpad} trains a \llm designed for program understanding by introducing the idea of a ``scratchpad" in which the \llm predicts the intermediate states of a program along with the final execution results. \cite{chen2022pot} extends the chain-of-thoughts~\cite{wei2022cot} prompting style in \nlp to propose program-of-thoughts where the prompt contains an explicit command to construct the program step-by-step. However, these work still generates a complete result (i.e., final program execution or code), albeit with intermediate results, in one shot, whereas our conversational \apr samples multiple times \llm{s} with different inputs to obtain one output plausible patch.  

Different from one-shot methods, \cite{austin2021synthesis} investigated the ability for \llm{s} to use human feedback in a conversational manner for program synthesis. The approach works by keeping a conversation of previously generated code and correcting any mistake using natural language feedback provided by human developers. \cite{Nijkamp2022CG} manually created a multi-step synthesis dataset where each target program is broken down into multiple smaller steps where only a few lines of code needs to be generated. They then sample the model multiple times to iteratively complete each smaller step and concatenate them together to form the final program. While these described techniques involve iteratively sampling from the model with new feedback similar to a conversational manner, our work can automatically create this feedback through testcase execution without any human-in-the-loop.

\section{Conversational \apr} 
\label{sec:approach}

We propose a conversational \apr approach to prompt \llm patch generation by combining previously generated patches and validation feedback in a conversational manner. Contrasting with the classic Generate and Validate (G\&V) \apr approach that first generates a large number of candidate patches and then validate each one to find a list of plausible patches, conversational \apr interleaves generation and validation to provide immediate feedback for the new candidate patch. Different from previous \apr tools which make use of \llm{s} through sampling given the same prefix/context for each bug, conversational \apr approach aims to incorporate feedback information after each generation (if the candidate patch failed to pass all tests) as new context for subsequent generations. Specifically, the feedback information includes both the incorrect generated patch and its associated failed testcase information.

\begin{figure}
    \includegraphics[width=\linewidth]{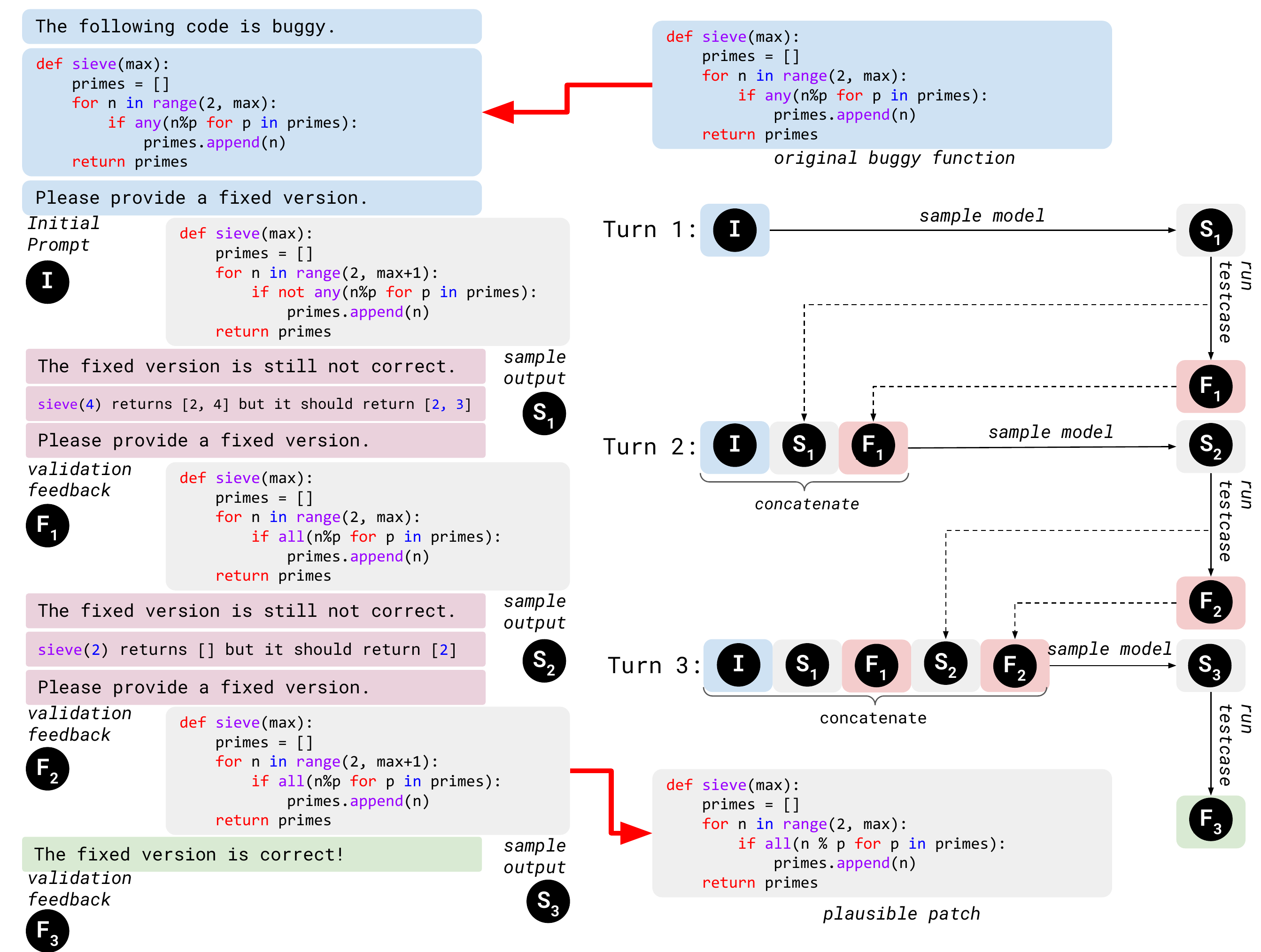}
    \centering
    \caption{Overview of conversational \apr with an illustrative example in fixing the buggy \CodeIn{sieve} function}
    \label{fig:overview}
\end{figure}

Conversational \apr involves iteratively obtaining new candidate patches from the \llm by using previously generated patches/validation results as feedback.\ We refer to this process as a \emph{turn}, where each turn includes three different steps: 1) construct new a prompt based on prior feedback 2) sample the model to produce a sample output function 3) validate the sample output function against testcases to obtain validation feedback. Multiple turns in sequence is defined as a \emph{chain}. The terminating conditions are that the sample output patch is able to pass all testcases (i.e., a plausible patch is obtained) or the maximum number of turns (length of the chain) is reached. Note that each turn (all three steps) are done automatically without needing any human-in-the-loop, this allows conversational \apr to be an automatic approach for program repair.

\subsection{pipeline \& example}

Figure~\ref{fig:overview} shows an illustrative example of a conversation chain (multiple turns) and an overview of the pipeline of the conversational \apr approach. We first take in as input the original buggy function and a set of testcases which contains some failing tests that expose the underlying bug. In the example, the buggy function (\CodeIn{sieve}) attempts to use to sieve algorithm to calculate the list of prime numbers below the integer input (\CodeIn{max}). The location of the bug occurs on line 4 where the buggy function incorrectly uses \CodeIn{any} instead of \CodeIn{all}. This bug is exposed by the testcase of \CodeIn{sieve(2) = [2]} where the buggy function incorrectly returns an empty array \CodeIn{[]}. 

\begin{itemize}[noitemsep, leftmargin=*]
    \item \textbf{Turn 1:} We first create an initial prompt \circled{$\text{I}$} using the original buggy function which contains natural language description to indicate that the function is buggy (\CodeIn{The following code is buggy}) and the task we want the \llm to solve (\CodeIn{Please provide a fixed version}). We then sample the model using the initial prompt \circled{I} to obtain the first sample output function \circled{$\text{S}_1$}. The change is made to line 4 where the function in \circled{$\text{S}_1$} negated the original \CodeIn{if} condition. We then validate \circled{$\text{S}_1$} against the list of tests and found that while the new patch is able to successfully pass the previous failing test of \CodeIn{sieve(2) = [2]}, it returns \CodeIn{[2, 4]} for \CodeIn{sieve(4)} when the correct output should be \CodeIn{[2, 3]}. This validation information \circled{$\text{F}_1$} is collected as feedback to use during the next conversation turn. 
    
    \item \textbf{Turn 2:} Different from turn 1, where the input to the \llm is just the initial prompt \circled{$\text{I}$}, now we provide the model also with the previously generated patch and its failing testcase. In short, we construct the validation feedback \circled{$\text{F}_1$} by using the failing testcase and indicate to the model that the previous sample \circled{$\text{S}_1$} is still not correct (\CodeIn{The fixed version is still not correct}) and the new task (\CodeIn{Please provide another fixed version}). We then concatenate the initial prompt, first sample output function and the validation feedback \{\circled{$\text{I}$}, \circled{$\text{S}_1$}, \circled{$\text{F}_1$}\} together as the input to the \llm. As such, the model is able to not only use the original buggy function but also use the previously generated sample and its testcase feedback to generate a new patched function. Similar to turn 1, we obtain \circled{$\text{S}_2$} and \circled{$\text{F}_2$} where the correct line 4 is obtained (switching \CodeIn{any} to \CodeIn{all}) but the candidate patch function incorrectly reduced the upper range of the for loop by 1.
    
    \item \textbf{Turn 3:} Similar to turn 2, we first construct the new validation feedback \circled{$\text{F}_2$} from the previous failing test case. We then concatenate all previously sampled output along with its validation feedback in sequence to produce \{\circled{$\text{I}$}, \circled{$\text{S}_1$}, \circled{$\text{F}_1$}, \circled{$\text{S}_2$}, \circled{$\text{F}_2$}\}. Using this input, we then sample the \llm again to produce the next candidate patch \circled{$\text{S}_3$}. We observe that this candidate patch correctly fixes the underlying bug and this is indicated by its validation \circled{$\text{F}_3$} where it is able to pass all the testcases. The program repair process is then terminated as we have obtained our plausible patch \circled{$\text{S}_3$}. 
\end{itemize}

Compared to prior approach in \apr based on \llm{s} which simply samples from a pre-defined prompt/context, conversational \apr leverages the previously missing key feedback information in the form of testcase results to prompt future patch generations. The testcase feedback not only tells the \llm that the previous patches are incorrect (i.e. leading to more unique patches) but also provides input and output examples which helps the model to understand the underlying functionality of the function (i.e. leading to more correct patches).   

\subsection{Design Decisions}

In the above example illustrated in Figure~\ref{fig:overview}, we show the overall pipeline of conversational \apr. However, there are different design decisions which can impact the performance of the approach:

\textbf{Prompt engineering.} Prompting has been shown to be an effective way of leveraging \llm{s} on various downstream tasks without needing any explicit fine-tuning. In conversational \apr approach, we follow the style of prior work~\cite{xia2022repairstudy} in providing a short and concise prompt with respect to the description of the input and the task we want to model to solve. Additionally, we follow prior guidelines and kept the prompt to be open-ended rather than to restrict the generation with a close-ended prompt. One particular important prompt constructing is validation feedback in providing the failing testcase to the \llm. In the Figure~\ref{fig:overview} example, we provide a \textit{functional} prompt that directly invokes the function and highlight the discrepancy between output and expected testcase output. We refer to this as functional prompt since it directly calls the function with input parameters similar to what one would do in code. In Section~\ref{sec:ablation}, we compare this style of validation prompting with other methods including without any testcase information to demonstrate the benefit of including validation feedback to the model. 

\textbf{Maximum chain length.} Recall that a conversation chain refers to the continuous sequence of turns to fix a bug. A chain is demonstrated in Figure~\ref{fig:overview} with a chain length of 3. Along with finding a plausible patch, a preset value for the maximum chain length is also a terminating condition since the \llm used will have a maximum context window and cannot take in arbitrary length inputs. Once this maximum chain length is reached, conversational \apr will restart from the beginning (i.e., by crafting initial prompt again) with a new chain conversation. The maximum chain length is a parameter which controls how much \textit{history} the \llm may receive. A maximum chain length of 1 refers to the base case of sampling from the initial prompt over and over again, meaning the model does not know any of the previously generated incorrect patches. A higher maximum chain length means the model can see multiple previously failed patches, however this also may not be beneficial as it can cause the \llm to repeat some of the earlier patches or get stuck on a particular implementation of the function. In Section~\ref{sec:ablation}, we evaluate the effect of the chain length has on repair performance.

\section{Datasets}

In this section, we describe the \llm{s} used in our evaluation and also the repair benchmark used to evaluate our proposed technique. 

\subsection{\llm{s}}

\begin{table}
\small
    \caption{Evaluation \llm overview}
    \centering
    \label{tab:model_stats}
    \begin{tabular}{@{}lrcc@{}}
    \toprule
    \textbf{Model} & \textbf{\#Parameters} & \textbf{Context Window} & \textbf{Training Strategy} \\
    \midrule
    \codegenmono  & 350M/2B/6B/16B & 2048 & Unsupervised CLM\\
    \codegenmulti & 350M/2B/6B/16B & 2048 & Unsupervised CLM\\
    \codex & 12B & 4096 & Unsupervised CLM\\
    \chatgpt & \(\sim\)175B & \(\sim\)4000 & \makecell{Reinforcement Learning\\from Human Feedback + CLM} \\
    \bottomrule
    \end{tabular}
\end{table}

In our work, we evaluate 10 different \llm{s} to not only demonstrate the effect of scaling behavior on our proposed conversational \apr approach but also to evaluate how different pre-training and model design contribute to the overall effectiveness. Table~\ref{tab:model_stats} presents an overview of the studied \llm{s}. Column \textbf{Model} is the model name, \textbf{\#Parameters} indicates the number of model parameters, \textbf{Context Window} represents the size of the context window, and \textbf{Training Strategy} refers to the training strategy used. 
\begin{itemize}[noitemsep,leftmargin=*]
    \item \textbf{\codegen~\cite{Nijkamp2022CG}.} A family of autoregressive \llm{s} trained using Causal Language Modeling (CLM) objective (next-token-prediction) ranging from 350M to 16B in parameter size. \codegen is first trained on the open-source ThePile~\cite{gao2020pile}, containing 22 diverse text-based datasets. The models are then trained on BigQuery~\cite{BigQuery}, a dataset of open-source code from 6 programming languages. We refer to these models (trained on ThePile then BigQuery) as \codegenmulti. \codegenmulti is then further trained on a dataset containing large amounts of Python GitHub code to produce \codegenmono. In our experiments, we use \codegenmono for repair benchmarks in Python and \codegenmulti for repair benchmarks in other programming languages by refer to them both as \codegen for simplicity. 
    \item \textbf{\codex~\cite{chen2021codex}.} A programming language focused autoregressive model based on the GPT-3 architecture~\cite{brown2020gpt3}. \codex is first initialized with GPT-3 weights from training on natural language corpus and then fine-tuned using next-token-prediction on a large dataset of code files. While \codex also contains a version which can take in suffix tokens (i.e., fill-in code in the middle), for our experiments, we only use \codex by providing the prefix context. 
    \item \textbf{\chatgpt~\cite{chatgpt}.} A conversational-based \llm first initialized from GPT-3.5 model and then fine-tuned using Reinforcement Learning from Human Feedback (RLHF)~\cite{ziegler2019rlhf}. \chatgpt is first fine-tuned based on supervised learning where human provides example responses to prompts in the dataset. Using this fine-tuned model, a reward model is then trained by sampling multiple outputs of the model from a given prompt and again using a human to rank the outputs. The reward model is used in the reinforcement learning step where Proximal Policy Optimization~\cite{schulman2017ppog} is used to fine-tune \chatgpt. Different from the \codex and \codegen, \chatgpt through the usage of RLHF and fine-tuning data is designed for conversation where the usage encourages a dialogue format. Note that much of the \chatgpt model detail is unknown to the public, therefore, we can only provide an approximate value for the number of parameters\footnote{As \chatgpt is fine-tuned on GPT-3.5, we assume a similar number of parameters as GPT-3.5} and context window size~\cite{chatgptcontext} according to verified sources. 
\end{itemize}

\subsection{Benchmarks}

\begin{figure}
    \includegraphics[width=\linewidth]{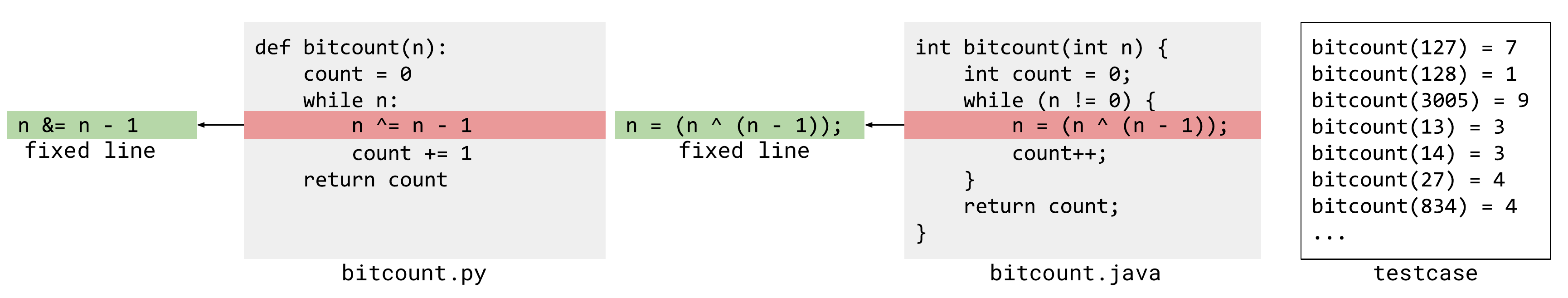}
    \centering
    \caption{Example bug in both Python and Java in \quix along with the testcases} 
    \label{fig:quixbugs_example}
\end{figure}

We use the \quix~\cite{lin2017quixbug} repair benchmark to evaluate our proposed conversational \apr approach. \quix has been widely used to evaluate many repair tools including both \learning~\cite{ye2022rewardrepair, zhu2021recoder, jiang2021cure, drain2021deepdebug} and \llm for \apr~\cite{xia2022alpharepair, xia2022repairstudy, kolak2022patch, prenner2021codexws} approaches. \quix dataset contains the same 40 bugs and it associated correct patch in both Python and Java. These bugs are self contained functions based on classic algorithms and it usually only takes a single line change to fix the underlying bug. Each bug comes with a set of testcases which the buggy function failed to pass and can be used to evaluate any candidate patch generated. Figure~\ref{fig:quixbugs_example} shows an example bug for the \CodeIn{bitcount} function in \quix for both Java and Python. The bug occurs inside the while loop where the code incorrectly uses the \CodeIn{\^} operator instead of \CodeIn{\&} operator. We also show the example testcases for \CodeIn{bitcount} where it contains example inputs and the expected outputs when evaluated using the function. 

Out of the 40 bugs in \quix, we further filter out 10 bugs which includes testcases that are difficult to represent with our validation feedback prompt. For example, testcases for \CodeIn{detect\_cycle} involves a graph as an input to the function. In total, we use 60 bugs (30 and 30 respectively for Java and Python) for our evaluation. 
\section{Experimental Setup}

In this section, we describe the key research questions that our evaluation seek to answer, the evaluation metrics used and also describe the implementation details. 

\subsection{Research Questions}

We aim to investigate the following research questions:
\begin{itemize}[leftmargin=*]
    \item \textbf{RQ1:} What is the effectiveness of applying conversational \apr? 
    \item \textbf{RQ2:} How do different components of conversational \apr effect performance?
\end{itemize}

In RQ1, we first compare the performance of conversational \apr with a baseline approach used in prior \llm for \apr work where the patches are generated by continuously sampling from the same initial prompt. We further evaluate both the scaling effective of \llm as we increase the size of the model and also investigate the difference in performance of different pre-training strategies (e.g., \chatgpt vs. \codex). In RQ2, we dive deeper into the different parameters of conversational \apr. Specifically, we evaluate how the length of the conversational chain and different validation feedback prompts affect the performance. 

\subsection{Evaluation Metrics} 

Our evaluation metric consist of the standard metric used to evaluate \apr tools: number of \emph{plausible patches}: patches which passes all the testcases and \emph{correct patches}: patches which are semantically equivalent to the reference developer patch. Additionally, since we are using sampling \llm{s}, we also define \emph{tries} as the number of samples needed to obtain a plausible/correct patch. This metric is useful when comparing two approaches/models that achieve similar number of bugs fixed, the one with fewer number of tries is preferred as we want to limit the number of times we have to sample the \llm. 

\subsection{Implementation} 

We implemented the \llm generation pipeline in Python using Hugging Face~\cite{HuggingFaceWebPage} implementation of the \codegen models. We access \codex through the OpenAI API by querying the \textit{code-davinci-002} engine. Since \chatgpt is not open-sourced and does not provide an official API endpoint (like \codex), we manually input the prompt and extract the outputs. For all models apart from \chatgpt, we use a default generation setting of nucleus sampling with top p = 0.95, temperature = 1, 50 samples per bug with a maximum chain length of 3. We generate and evaluate patches on a 32-Core workstation with AMD Ryzen Threadripper PRO 5975WX CPU, 256 GB RAM and 3 NVIDIA GeForce RTX 3090 GPUs, running Ubuntu 22.04.1 LTS. 
\section{Results}

\subsection{RQ1: Conversational \apr effectiveness}

\begin{table}
\small
\caption{Conversational \apr performance on both \quix-Python and \quix-Java compared with baseline sampling method. \#c/\#p refers to the number of correct / plausible patches.}
        \centering
        \label{tab:quixbugs}
        \begin{tabular}{@{}l | rr rr | rr rr@{}}
        \toprule
        \textbf{Models} & \multicolumn{4}{c}{\textbf{\quix-Python} }& \multicolumn{4}{c}{\textbf{\quix-Java}} \\
        & \multicolumn{2}{c}{\textbf{Sampling}} & \multicolumn{2}{c}{\textbf{Conversational}}  & \multicolumn{2}{c}{\textbf{Sampling}} & \multicolumn{2}{c}{\textbf{Conversational}} \\
        & \textbf{\#c/\#p} & \textbf{\#tries}  & \textbf{\#c/\#p} & \textbf{\#tries} & \textbf{\#c/\#p} & \textbf{\#tries}  & \textbf{\#c/\#p} & \textbf{\#tries}\\
        \midrule
        \codegen-350M & 7 / 10 & 20.5 & 8 / 11 & 18.4 & 4 / 4 & 24.2 & 5 / 5 & 23.5 \\
        \codegen-2B & 22 / 23 & 16.6  & 25 / 26 & 14.3 & 12 / 14 & 18.8  & 15 / 16 & 16.4\\
        \codegen-6B & 22 / 24 & 14.0  & 27 / 28 & 12.1 & 18 / 20 & 19.8 & 22 / 22 & 13.5 \\
        \codegen-16B & 29 / 29 & 5.6  & 30 / 30 & 4.8 & 24 / 25 & 14.5 & 28 / 29 & 13.2\\
        \codex & 29 / 30 & 4.6 & 30 / 30 & 3.8 & 28 / 30 & 7.2 & 29 / 30 & 5.7 \\
        \bottomrule
        \end{tabular}
\end{table}

\begin{table}
\small
\caption{\chatgpt and \codex comparison on \quix-Python and \quix-Java where each cell indicates the number of correct / plausible patches}
        \centering
        \label{tab:chatgptcompare}
        \begin{tabular}{@{}l rrr rrr@{}}
        \toprule
        \textbf{Models} & \multicolumn{3}{c}{\textbf{\quix-Python}} & \multicolumn{3}{c}{\textbf{\quix-Java}} \\
        \cmidrule(lr){2-4}\cmidrule(l){5-7}
        & \textbf{one try} & \textbf{two tries}  & \textbf{three tries} & \textbf{one try} & \textbf{two tries}  & \textbf{three tries} \\
        \midrule
        \codex & 16 / 16 & 21 / 21 & 24 / 24 & 11 / 12 & 18 / 19 & 21 / 22\\ 
        \chatgpt & 24 / 24 & 27 / 28 & 28 / 29 & 24 / 24 & 26 / 26 & 26 / 26\\ 
        \bottomrule
        \end{tabular}
\end{table}

We first evaluate the effectiveness of applying conversational \apr using validation feedback compared to prior method of sampling given the same prompt without any feedback. Table~\ref{tab:quixbugs} shows the results on \quix-Python and \quix-Java. We observe that by \emph{applying our feedback driven conversational \apr, we are able to improve the \# of correct and plausible patches for all unsupervisedly trained \llm across all model sizes}. Additionally, conversational \apr is also able to decrease the \# of tries (\# of samples) needed before obtaining the first plausible/correct patch. Compared to traditional sampling method of producing patches, conversational \apr is able to leverage the model's understanding of natural language feedback to indicate why the patch is incorrect. \llm{s} can use this validation feedback information to generate new patches that try to pass the previously failed testcase. Furthermore, conversational \apr also helps to reduce the number of repeated patches from sampling using the same prompt over and over again. By using the large context size of many state-of-the-art \llm{s}, conversational \apr can use recently generated incorrect patches as previous context to prompt the model to generate a new patch that is different.

\textbf{\chatgpt evaluation.} We now evaluate the performance of \chatgpt when using conversational \apr. Due to the requirement of manually inputting and extracting outputs from \chatgpt, we only use a single conversation chain with at most 3 tries (i.e. maximum chain length of 3). We compare with the best performing \llm of \codex from previous results under the same setting in Table~\ref{tab:chatgptcompare}. We observe that \emph{compared to \codex, which is trained in an unsupervised manner, \chatgpt which is fine-tuned using Reinforcement Learning from Human Feedback (RLHF) performed much better across the two repair datasets}. This improvement in result can be partially attributed to increase in model parameter size, but we believe this is also due to the dialogue-based fine-tuning dataset used in \chatgpt. Conversational \apr relies on the model understanding the validation feedback to condition the future generation in trying to generate a patch that passes the testcase. A more dialogue-oriented model such as \chatgpt is well suited for this task as both the training data and algorithm contain feedback driven loops. As \chatgpt and other dialogue-based \llm{s} become more popular, we believe conversational \apr can also be further improved through more usage of these \llm{s}. 

\subsection{RQ2: Component Analysis}
\label{sec:ablation}
\begin{figure}
    \includegraphics[width=\linewidth]{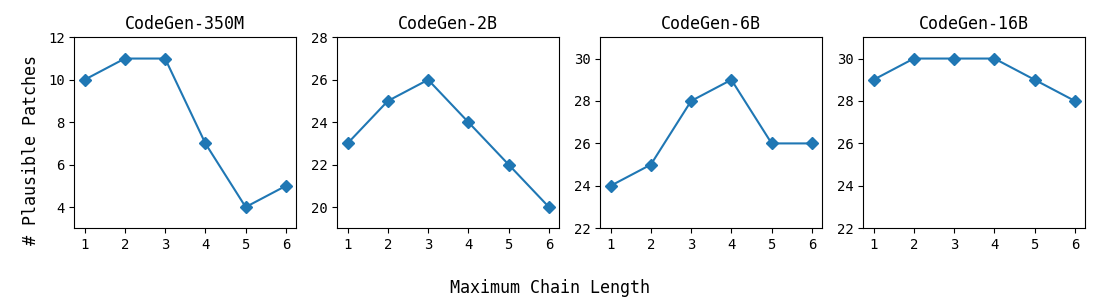}
    \centering
    \caption{Number of plausible patches for the 4 different \codegen models as we vary the maximum chain length on \quix-Python}
    \label{fig:chainlength}
\end{figure}

\begin{table}
\small
\caption{Prompting Style Evaluation on \quix-Python with each cell showing the number of plausible patches}
        \centering
        \label{tab:promptstyle}
        \begin{tabular}{@{}l rrr@{}}
        \toprule
        \textbf{Models} & \textbf{no testcase} & \textbf{natural language} & \textbf{functional} \\
        \midrule
        \codegen-350M & 9 & 11 & 11 \\
        \codegen-2B & 20 & 25 & 26 \\
        \codegen-6B & 24 & 27 & 28 \\
        \codegen-16B & 27 & 30 & 30 \\
        \codex & 29 & 30 & 30\\
        \bottomrule
        \end{tabular}
\end{table}

\textbf{Maximum chain length.} We first investigate the effect of different maximum chain length has on the repair performance. Figure~\ref{fig:chainlength} shows the number of plausible patches when we vary the maximum chain length from 1 to 6 for the 4 \codegen models. Recall from Section~\ref{sec:approach} that chain length refers to the number of turns (each turn consist of generating and validating a new patch) in a conversation chain. A maximum chain length of 1 is the simple sampling from the same initial prompt baseline (used in prior \llm for \apr tools). As we increase chain length, the model has to take in more and more previous context in the form of prior generations and feedbacks. We observe that the performance increase as we start from a small chain length and reaches the maximum around 3 or 4 and then decrease as chain length continue to increase. The decrease in number of plausible patches once we reach a high chain length is because the context may be too much for the model to handle since it can include multiple previously failed patches. We also observe that this decrease is more significant in smaller models, where larger models are less affected by longer chain length, showing the ability for larger models to better capture the long term context dependencies. This shows that the optimal chain length to use for conversational \apr can be dependent on the individual \llm used. 

\textbf{Feedback prompting style.} We now evaluate the effect of the feedback prompting style used in our conversational \apr. Table~\ref{tab:promptstyle} shows the number of plausible patches using different validation prompts in \quix-Python. Column \textbf{no testcase} does not include any testcase feedback (only states that the patch is not correct), \textbf{natural language} describes the failing testcase (e.g., \CodeIn{when input is 2, the patch incorrectly returns [] but it should return [2]}) and \textbf{functional} which is the default prompting style discussed in Section~\ref{sec:approach}. We observe that different prompting style does have an effect on the final performance of conversational \apr. Starting from no testcase prompt, we can improve performance by adding specific testcase feedback information on top of telling the \llm that the patch is not correct. We also observe that the functional prompting style, using the buggy/patch function name and passing parameters (see Figure~\ref{fig:overview}), performs the best. Functional prompting style conveys the failing testcase information in a more concise and natural way by phrasing the testcase input and expected output relationship as a function call. 
\section{Conclusion}

We propose conversational \apr, a new paradigm for program repair that interleaves patch generation with validation to provide immediate feedback for \llm{s} to better prompt future generated patches. Compared to previous \llm for \apr approaches that only sample from the same input, conversational \apr iteratively builds the input by concatenating previously incorrect patches and validation feedback. This allows for the model to avoid generating previously incorrect patches and also understand the semantic meaning of the function through validation feedback. Our evaluation on 10 different \llm{s} shows the improvement of conversational \apr over the baseline sampling method used in prior \llm for \apr tools. Furthermore, we demonstrate the promising future of applying \chatgpt, a conversational/dialogue driven \llm, for conversational \apr, or \apr in general for the first time.

\bibliography{main}
\bibliographystyle{iclr2023_conference}

\end{document}